\documentclass[aps,showpacs,prl,toolkits,twocolumn]{revtex4}
\usepackage{epsfig,amssymb,amsfonts}

\makeatletter
\renewcommand{\@makefntext}[1]{\parindent=1em\noindent\hbox to 1.8em{\hss$^{\@thefnmark}$}#1}
\renewcommand{\@footnotemark}{\hbox{\mathsurround=0pt$^{\@thefnmark}$}}

\makeatother

\begin{document}
\title{A simple toy model for effective restoration of chiral
symmetry in excited hadrons.}
\author{Thomas D.~Cohen${}^1$   and Leonid Ya. Glozman${}^2$}
\affiliation{${}^1$Department of Physics, University of Maryland,
College Park, MD 20742, USA} \affiliation{${}^2$Institute for
Theoretical Physics, University of Graz, Universit\"atsplatz 5,
A-8010 Graz, Austria}

\begin{abstract}
A simple solvable toy model exhibiting effective restoration
of chiral symmetry in excited hadrons is constructed. A salient 
feature is that while physics of the low-lying states is
crucially determined by the spontaneous breaking of chiral
symmetry, in the high-lying states the effects of chiral
symmetry breaking represent only a small correction. Asymptotically
the states approach the regime where their properties are determined
by the underlying unbroken chiral symmetry.
\end{abstract}

\maketitle

One striking feature of the hadron spectrum is the presence of
nearly degenerate resonances of opposite parity relatively high in
the excitation spectrum. It has been argued that this phenomenon
might be a manifestation of an ``effective restoration of chiral
symmetry'' \cite{G1,CG1,CG11,G2,G22,G3,G4,G44}.  The central idea underlying
this notion is the possibility that the coupling of these highly
excited states to the dynamics driving spontaneous chiral symmetry
breaking could get progressively weaker for progressively massive
hadrons leading to the excited states being quite insensitive to
the effects of spontaneous chiral symmetry breaking and hence act
qualitatively very much as if they were in the Wigner-Weyl mode.
In such a situation the resonances would be expected to form
approximate linearly realized chiral multiplets whose members were
nearly degenerate.  Such a scenario naturally predicts multiplets
of states with opposite parities. That the linear realization of
the chiral symmetry could be relevant to the low-lying hadrons,
where, however, the chiral symmetry is strongly broken, have been
explored in the context of a certain class of models \cite{Jido,Jido1}
and in heavy-light quark physics \cite{HQ,HQ1,HQ11,HQ111}.

It is important to precisely characterize what is implied under
effective restoration in \cite {G1,CG1,CG11,G2,G22,G3,G4,G44}, because 
sometimes
it is erroneously interpreted in the sense that the highly-excited
hadrons are in the Wigner-Weyl mode. The mode of symmetry is defined
only by the properties of the vacuum. If symmetry is spontaneously
 broken in the vacuum, then it is the Nambu-Goldstone mode and the
 {\it whole} spectrum of excitations on the top of the vacuum is in
 the Nambu-Goldstone mode. However, it may happen that the role
 of the chiral symmetry breaking condensates of the vacuum becomes
 progressively irrelevant in excited states. This means that the
 chiral symmetry breaking effects (dynamics) become less and less
 important in the highly excited hadrons and asymptotically 
 the states approach the regime where their properties are determined
 by the underlying unbroken chiral symmetry (i.e. by the symmetry in
 the Wigner-Weyl mode).

Although the basic idea has been discussed in the past,
it is useful to construct a simple model which
illustrates clearly the basic idea. Such 
a model has the virtue of demonstrating explicitly that the
idea is consistent with the underlying concepts of chiral
symmetry and spontaneous chiral symmetry breaking. This may be
of use in preventing confusion 
about the physical content of "effective chiral restoration".
 We
will then use the behavior in this model to clarify the physical
meaning of the given phenomenon.

Consider as an example the following model which, while having no
particular physical significance illustrates quite clearly 
the physical content of effective chiral restoration of the
type discussed in refs. \cite {G1,CG1,CG11,G2,G22,G3,G4,G44}.  
The model contains
an infinite number of $\pi$ and $\sigma$ mesons.  In this respect
the model mimics large $N_c$ QCD. We denote the $j^{\rm th}$ pion
($\sigma$ meson) $\pi_j$ ($\sigma_j$). These fields enter the
Lagrangian in a chirally invariant way as members of $\left (
\frac{1}{2},\frac{1}{2} \right ) $ chiral multiplets:
\begin{eqnarray}\left [V^a,\pi^b_j \right]  =  
i \,\epsilon^{a b c} \, \pi^{c}_j
&{}& \left [ V^a,\sigma_j \right ]  =  0 \nonumber\\
\left[A^a,\pi^b_j \right] = i \delta_{a b} \, \sigma_j \; &{}&
\left[A^a,\sigma_j \right] = i \pi^a_j
\end{eqnarray}
where $V^a$  ($A^a$) represent the generators of vector (axial)
rotations. To simplify the analysis by reducing the number of
possible couplings,  in addition to chiral symmetry the model
has an infinite number of discrete symmetries: it is invariant
under $\sigma_j\rightarrow -\sigma_j$ for all $j$.  The discrete
symmetries ensure that each type of field always enters the
Lagrangian in even powers. The Lagrangian is given by
\begin{widetext}
\begin{eqnarray}
{\cal L} &=& \sum_j \frac{1}{2} \left ( \partial^\mu  \sigma_j \,
\partial_\mu \sigma_j \, +
\partial^\mu  \vec{\pi}_j \cdot \partial_\mu \vec{\pi}_j  \right )
 - \frac{m_o^2}{2} \left ( \alpha (\sigma_1^2 + \vec{\pi}_1 \cdot
\vec{\pi}_1 )  +  \frac{g}{2 m_o^2} (\sigma_1^2 + \vec{\pi}_1
\cdot \vec{\pi}_1 )^2 \right ) \nonumber \\ & - &  \frac{m_o^2}{2}
\sum_{j=2}^\infty \, \left( j^2 (\sigma_j^2 \, + \vec{\pi_j} \cdot
\vec{\pi_j}) \, + \, \frac{g}{j \, m_o^2 \,} \left ( (\sigma_1
\sigma_j + \vec{\pi}_1\cdot \vec{\pi}_j)^2 \, + \, (\sigma_1^2  +
\vec{\pi}_1\cdot \vec{\pi}_1) \, ( \sigma_j^2 + \vec{\pi}_j\cdot
\vec{\pi}_j)  \right )\right) + g V_4\label{L}
\end{eqnarray}
\end{widetext}
where $m_o$ has dimensions of mass and $\alpha$ and $g$ are
dimensionless constants and $V_4$ is some function of the fields
$\sigma_j, \vec{\pi_j}; ~ j>1$  consistent with the symmetries
and with the potential whose terms are quartic in the fields. The
model is chosen so that $j=1$ fields play a special role in the
chiral broken phase: $\sigma_1$ (and no other fields) acquires a
vacuum expectation value and the excitation associated with
$\pi_1$ becomes massless. The parameter $\alpha$ controls
spontaneous symmetry braking; $\alpha>0$ yields the Wigner-Weyl
mode while $\alpha<0$ yields the Nambu-Goldstone mode.  As will be
seen below the analysis is  independent of the particular form
picked for $V_4$.

The interaction terms in the model of Eq.~(\ref{L}) are controlled
by the parameter $g$. For $g \ll 1$, the theory is weakly coupled
and hence can be treated classically. The Lagrangian is
parameterized in such a way that the dependence of the mass
spectrum on $g$ only arises through loop contributions which can
be neglected in the weak-coupling limit. Note, however, that even
in the weak coupled limit the interaction terms play an essential
role when $\alpha < 0$ since it determines the amount of chiral
symmetry breaking.  We consider here the weak coupling limit of
the theory which is analytically tractable---indeed trivial---but
is a perfectly legitimate chiral theory.  We will study the theory
in the weakly coupled regime where it is tractable.

\begin{figure}[t]
\includegraphics{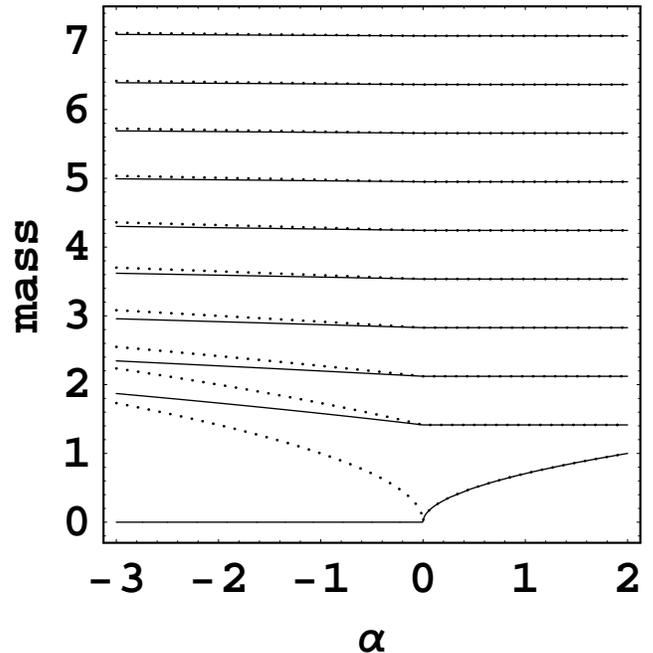}
\caption{The mass spectrum of the model in Eq.~(\ref{L}) in units
of the mass parameter $m_o$.  The solid lines correspond to pions
while the dotted lines correspond to $\sigma$-mesons. \label{toy}}
 \end{figure}

If one imposes isospin invariance then it is easy to see that the
minimum of the potential is given by:
\begin{eqnarray}
\langle \sigma_j \rangle & =&  0 \; \; \; \; {\rm for} \, \, \alpha > 0 \nonumber \\
\langle \sigma_j \rangle & = & \pm \delta_{j 1} \, m_o \,
\sqrt{\frac{-\alpha}{g}} \; \; \; \; {\rm for} \, \, \alpha \le 0
\; .\end{eqnarray} By expanding quadratically around the minimum
of the potential one can find the mass spectrum.
\begin{equation}
 {\rm for}\, \,  \alpha > 0   \; \left \{ \begin{array}{l}
m^2_{\pi_1} =  \alpha m_o^2  \\ \\
m^2_{\sigma_1}  = \alpha m_o^2  \\ \\
m^2_{\pi_j}  = j^2 m_o^2 \; \; (j \ge 2) \\ \\
m^2_{\sigma_j}  =  j^2 m_o^2\; \; (j \ge 2)
\end{array} \right . \nonumber
\end{equation}
\begin{equation}
{\rm for} \, \,  \alpha \le 0 \; \left \{ \begin{array}{l}
m^2_{\pi_1} =  0 \\ \\
m^2_{\sigma_1}  = - 2 \alpha m_o^2  \\ \\
m^2_{\pi_j}  =  j^2  m_o^2+ \frac{2 g\langle \sigma_1 \rangle^2
}{j}
\\\; \; \; \; \; =\left( j^2 + \frac{2 \alpha}{j} \right ) m_o^2\; \; (j \ge 2) \\ \\
m^2_{\sigma_j}  =   j^2 m_o^2 + \frac{4  g \langle \sigma_1
\rangle^2 }{j}
\\\; \; \; \; \; =  \left ( j^2 + \frac{4 \alpha}{j} \right)  m_o^2 \; \; (j \ge 2)
\end{array} \right .
\end{equation}
This procedure is legitimate since we are working in the weak
coupling limit and quantum (loop) corrections are suppressed.  The
spectrum is shown in Fig.~\ref{toy}.
It is apparent from Fig.~\ref{toy}, that this model in the
spontaneously broken phase exhibits the phenomenon of effective
chiral restoration.  Consider, for example, the region near
$\alpha=-1$. While the lowest-lying states have no hint of a
chiral multiplet structure, as one goes higher in the spectrum the
states fall into nearly degenerate multiplets which to
increasingly good approximation look like  pions and
$\sigma$-meson in linearly realized $\left (
\frac{1}{2},\frac{1}{2} \right )$ representations.

Suppose that there is a "no-go theorem" forbidding effective chiral restoration
in excited hadrons  and the  existence of approximate parity
doublets in the Nambu-Goldstone phase cannot be a manifestation of
chiral symmetry.  If this were correct then the appearance of
approximate parity doublets in the Nambu-Goldstone phase of this
model must be unrelated to chiral symmetry. However, this is
obviously not the case: the near degeneracy of the  parity
doublets reflects the underlying chiral symmetry of the model. The
key point is that high-lying states are very insensitive to the
chiral symmetry breaking order parameter $\sqrt{g}
 \langle \sigma_1 \rangle$:
\begin{eqnarray}
\frac{\partial m_{\sigma_j}}{\partial  \left(\sqrt{g} \langle
\sigma_1 \rangle \right) } & = & \frac{4 m_o \sqrt{-\alpha}}{ j
m_{\sigma_j}}
\rightarrow \frac{4 \sqrt{-\alpha} }{ j^2 } \nonumber \\
\frac{\partial m_{\pi_j}}{\partial \left ( \sqrt{g} \langle
\sigma_1 \rangle \right ) } & = & \frac{2 m_o \sqrt{-\alpha}}{ j
m_{\pi_j}} \rightarrow \frac{2 \sqrt{-\alpha}}{ j^2 }
\end{eqnarray}
where the arrow indicates the asymptotic behavior at large $j$.
Clearly, at large $j$ the masses become increasingly insensitive
to the chiral order parameter and the spectrum approximates a
Wigner-Weyl mode spectrum increasingly accurately.

The scenario in which high mass resonances become increasingly
insensitive to the chiral order parameter and hence form into
approximate chiral multiplets in the spectrum is precisely what was meant
by effective chiral restoration in refs.~\cite{G1,CG1,CG11,G2,G22,G3,G4,G44}.
Thus, since there is at least one chirally invariant  model in
which this happens, there cannot be any
``no-go theorem'' for effective chiral restoration.

One might legitimately argue that the model in Eq.~(\ref{L}) is
highly artificial.  The model was designed---with malice
aforethought---to ensure effective chiral restoration.  This was
done in part by fixing the term in the Lagrangian which couples to
the chiral order parameter to scale like $1/j$ and hence become
weak for large $j$.  This is, of course, true. However, the point
of the exercise was simply to demonstrate that there is a solvable
model that exhibits
effective chiral restoration and this model helps to clarify a physical
meaning of this phenomenon.
However artificial the model, it is adequate for this
purpose.

The model above was formulated in terms of the fields that
transform linearly under the chiral group. Then a question arises
whether this model illustrates some generic behavior or it is
only specific to the linear realization of the chiral symmetry?
Indeed,
in the Nambu-Goldstone mode one can always  make a field
redefinition to the standard nonlinear realization \cite{C,C1}
 in which fields
of opposite parity are decoupled under,
{\it i.e.} do not transform into each other under chiral
transformation. In this case the act of making an axial rotation
does not transform a field into its chiral partner but instead
creates a massless Goldstone boson (pion) from the vacuum.
However, such a redefinition is
unphysical and the Lagrangian of Eq.~(\ref{L}) rewritten in terms
of these new fields cannot alter the spectrum.
This reflects a general situation that the field redefinitions
themselves cannot modify any physical content of the theory. In
this context it is useful to recall that the physics is not in the
fields, but in the states, which appear once one applies fields on
the vacuum. The physics of these states is controlled only by the
microscopical theory.
Clearly the chiral
symmetry can impose no constraints on these decoupled fields  in
the spontaneously broken phase.  However this does not rule
out effective chiral restoration. Generically, in the
Nambu-Goldstone mode, the properties of states are sensitive to
the dynamics of chiral symmetry breaking---this is why  the states
generically do not form  {\it exact} chiral multiplets.

If however, there exist states in the spectrum which for some
reason are insensitive or weakly sensitive
to the dynamics of chiral symmetry
breaking, there will be approximate relations between the
properties of the levels dictated by the underlying chiral symmetry
(in the limit of complete insensitivity
the states must look identical to exact linear chiral multiplets).  The
reason for this is simple: one
can reverse the transformation from the standard nonlinear
realization and rewrite things in terms of a set of linear
realized fields; theory in terms of these fields is equally
physical as the standard nonlinear realization. These fields
correspond to states which are related to each other under chiral
symmetry except due to the interactions with the chiral order
parameter and to the Goldstone bosons. As these interactions
become small approximate chiral multiplets emerge. This is
explicitly seen in our toy model since the high-lying states do
decouple from the vacuum condensate and from the Goldstone bosons.
Indeed, it was argued in ref. \cite{G3,N} that the high-lying
hadrons do decouple from the Goldstone bosons once they decouple
from the quark condensate.

Thus, the issue is a quantitative one and not merely a qualitative
one. In the Nambu-Goldstone mode the states should be
sensitive to the dynamics of chiral symmetry breaking.  
  The question is how much.  As
demonstrated above, it is certainly possible in some models for
high lying states to have very small coupling to the dynamics of
symmetry breaking and, hence, to very good approximation to fall
into chiral multiplets. The conjecture of effective chiral
restoration in hadronic physics is that QCD behaves in the same
manner as this simple model.

As noted above the model in Eq.~(\ref{L}) is highly artificial.
However, it does illustrate many of the salient
points relevant to the issue.
Firstly, it illustrates the most important point: while the physics of the
low-lying states is crucially determined by the spontaneous breaking
of chiral symmetry, in the high-lying states the effects of
chiral symmetry breaking represent only a small correction.
Secondly, it shows the
gradual nature of the conjectured effect.  The effect is never
absolute but always approximate; for any given strength of the
coupling $\alpha$ it becomes increasingly accurate as one goes up
in the spectrum. Thirdly, it makes very clear that the key issue
is the coupling of the state to the dynamics responsible for
spontaneous chiral symmetry breaking---in this case the coupling
to $\sqrt{g} \langle \sigma_1 \rangle$ which plays the role of the chiral order
parameter.

Although we have constructed a simple tractable toy model
explicitly exhibiting the effective restoration of chiral symmetry
in excited hadrons, the essential question of whether the parity partners
seen in excited hadrons are due to effective chiral restoration
remains open.  As discussed in \cite{CG1,G4,G44}, there are  general
arguments leading to the expectation that as one goes to
asymptotically high mass states the sensitivity to the chiral
condensate decreases. This in turn leads to the expectation of
effective chiral restoration in the hadron spectrum, {\it
provided} that discernable hadrons still exist in the regime where
the sensitivity to the chiral condensate becomes negligible.
However, at present there is no reliable theoretical tool which
allows one to answer the question of whether discernable hadronic
resonances persist high enough in the spectrum to reach the regime
of effective chiral restoration or whether the spectrum
essentially melts into the QCD continuum before this point. There
is an interesting theoretical limit---the large $N_c$ limit---in
which the meson spectrum is infinite and remains discrete and the
phenomenon of effective chiral restoration ought to occur. 
Some discussions of the rate of symmetry restoration can be found
in \cite{SHIFMAN,GOLTERMAN}. The results of the solvable model
of the 't Hooft type in 3+1 dimensions are presented in \cite{W}.
While
the large $N_c$ argument is interesting theoretically and shows
how the phenomenon may come about, it does not provide a
compelling theoretical argument for the $N_c=3$ world. Similarly,
the present empirical evidence is  not compelling but we take it
to be suggestive.

There is an empirical way to verify the idea, however. If the
parity doubling observed at present is indeed due to chiral
restoration, then some of the missing states in the chiral
multiplets with approximately known masses should be
experimentally found. This point is a legitimate subject for
discussion.

\begin{acknowledgments}

Correspondence with R. Jaffe, D. Pirjol, A. Scardicchio,
 M. Shifman and A. Vainstein is acknowledged.  
LYaG acknowledges the support from the
the Austrian Science Fund, projects
P16823-N08 and P19168-N16. TDC acknowledges
the support of the United States Department of Energy.

\end{acknowledgments}


\begin{thebibliography}{99}

\bibitem{G1} L. Ya. Glozman, Phys. Lett. B {\bf 475}, 329 (2000).
 \bibitem{CG1} T. D. Cohen and L. Ya. Glozman, Phys. Rev. D {\bf 65}, 016006
 (2002).
 \bibitem{CG11} T. D. Cohen and L. Ya. Glozman, 
 Int. J. Mod. Phys. A {\bf 17}, 1327 (2002).
\bibitem{G2} L. Ya. Glozman, Phys. Lett. B {\bf 539}, 257 (2002).
\bibitem{G22} L. Ya. Glozman, Phys. Lett. B {\bf 587}, 69 (2004).
\bibitem{G3} L. Ya. Glozman, Phys. Lett. B {\bf 541}, 115 (2002).
\bibitem{G4}  L. Ya. Glozman, Int. J. Mod. Phys. A., {\bf 21}, 475 (2006).
\bibitem{G44}  L. Ya. Glozman, A. V. Nefediev, J.E.F.T. Ribeiro, Phys. Rev. 
{\bf D 72}, 094002 (2005).
\bibitem{Jido}D. Jido, T. Hatsuda, T. Kunihiro, Phys. Rev. Lett. ,{\bf 84}, 3252
(2000).
\bibitem{Jido1}D. Jido, M. Oka, A. Hosaka, Progr. Theor. Phys. {\bf 106},
873 (2001).
\bibitem{HQ} M. A. Nowak, M. Rho and I. Zahed, Phys. Rev. {\bf D
48}4370 (1993).
\bibitem{HQ1} M. A. Nowak, M. Rho and I. Zahed, Acta. Phys. Polon. B 
{\bf 35} 2377 (2004).
\bibitem{HQ11}  W. A. Bardeen and C. T. Hill, Phys. Rev. {\bf D49}, 409 (1994).
\bibitem{HQ111}  W. A. Bardeen E. J. Eichten and C. T. Hill, Phys. Rev.{\bf D49}, 409
(1994).
\bibitem{C} S. R. Coleman, J. Wess and B. Zumino, Phys. Rev. {\bf 177},
2239 (1969).
\bibitem{C1}  C. C. Callan, S. R. Coleman, J. Wess and B. Zumino,
Phys. Rev. {\bf 177}, 2247 (1969).
\bibitem{SHIFMAN} M. Shifman, hep-ph/0507246.
\bibitem{GOLTERMAN} O. Cata, M. Golterman, S. Peris, hep-ph/0602194.
\bibitem{W} R. F. Wagenbrunn, L. Ya. Glozman, hep-ph/0605247.
\bibitem{N}  L. Ya. Glozman, A. V. Nefediev, Phys. Rev. 
{\bf D 73}, 074018 (2006).

\end{thebibliography}
\end{document}